\documentclass[aps,prl,preprint,superscriptaddress]{revtex4}

\usepackage{times}

\DeclareFontFamily{OT1}{times}{}
\DeclareFontShape{OT1}{times}{m}{n }{ <-> ptmr }{}
\DeclareFontShape {OT1}{times}{bx}{n }{ <-> ptmb }{}
\DeclareFontShape {OT1}{times}{m }{it}{ <-> ptmri}{}
\DeclareFontShape {OT1}{times}{bx}{it}{ <-> ptmbi}{}
\usepackage{amsmath}
\usepackage{amsfonts}
\usepackage{amssymb}
\usepackage{latexsym}
\usepackage{graphicx}
\usepackage{setspace}

\usepackage{dcolumn}
\usepackage{bm}

\begin{document}

\preprint{Los Alamos National Laboratory LA-UR-08-0753}

\newcommand{\mba}{\ensuremath{\mathbf{a}}}
\newcommand{\mbb}{\ensuremath{\mathbf{b}}}
\newcommand{\mbc}{\ensuremath{\mathbf{c}}}

\title{\Large{Determinants of bistability in\\ induction of the
\textit{Escherichia coli lac} operon}}



\author{David W. Dreisigmeyer}
\affiliation{High Performance Computing Division}
\author{Jelena Stajic}
\thanks{Present address: Center for Cell Analysis and Modeling, University of Connecticut Health Center, Farmington, CT, 06030, USA.}
\affiliation{Center for Nonlinear Studies}
\author{Ilya Nemenman}
\affiliation{Computer, Computational, and Statistical Sciences Division}
\affiliation{Center for Nonlinear Studies}
\author{William S. Hlavacek}
\affiliation{Theoretical Division}
\affiliation{Center for Nonlinear Studies}
\author{Michael E. Wall}
\email[Corresponding author: ]{mewall@lanl.gov}
\affiliation{Computer, Computational, and Statistical Sciences Division}
\affiliation{Bioscience Division, Los Alamos National Laboratory, Los Alamos, NM 87545, USA}
\affiliation{Center for Nonlinear Studies}

\date{\today}

\begin{abstract}

We have developed a mathematical model of regulation of expression of
the {\em Escherichia coli lac} operon, and have investigated
bistability in its steady-state induction behavior in the absence of
external glucose. Numerical analysis of equations describing
regulation by artificial inducers revealed two natural bistability
parameters that can be used to control the range of inducer
concentrations over which the model exhibits bistability. By tuning
these bistability parameters, we found a family of biophysically
reasonable systems that are consistent with an experimentally
determined bistable region for induction by thio-methylgalactoside
(Ozbudak et al. Nature 427:737, 2004). The model predicts that
bistability can be abolished when passive transport or permease export
becomes sufficiently large; the former case is especially relevant to
induction by isopropyl-$\beta$, D-thiogalactopyranoside. To model
regulation by lactose, we developed similar equations in which
allolactose, a metabolic intermediate in lactose metabolism and a
natural inducer of {\em lac}, is the inducer. For biophysically
reasonable parameter values, these equations yield no bistability in
response to induction by lactose; however, systems with an
unphysically small permease-dependent export effect can exhibit small
amounts of bistability for limited ranges of parameter values. These
results cast doubt on the relevance of bistability in the {\em lac}
operon within the natural context of {\em E. coli}, and help shed
light on the controversy among existing theoretical studies that
address this issue. The results also suggest an experimental approach
to address the relevance of bistability in the {\em lac} operon within
the natural context of {\em E. coli}.

\end{abstract}

\pacs{}

\maketitle


\section{Introduction}
\label{SEC-INTRO}

In 1957, Novick and Weiner discovered that {\em Escherichia coli} can
exhibit discontinuous switching in expression of the {\em lac} operon,
with some cells expressing a large amount of $\beta$-galactosidase
($\beta$-gal), other cells expressing a small amount, and an
insignificant number of cells expressing an intermediate
amount~\cite{Novick57}. Recently, this effect was further
characterized using single-cell assays of fluorescence levels in a
population of {\em E. coli} cells carrying a {\em lac::gfp}
reporter~\cite{Ozbudak04}. The population exhibited a bimodal
distribution, with induced cells having over 100 times the
fluorescence level of uninduced cells. These observations have been
attributed to the existence of two steady states, i.e., bistability,
in the induction of {\em lac} in {\em E. coli}.

Recent modeling studies have emphasized the importance of determining
whether bistability in expression of {\em lac} is relevant within a
natural context
\cite{Savageau01,Savageau02,Yildrim03,Yildrim04,vanHoek06}. This
question remains open because experimental studies have focused on the
response of {\em lac} expression to artificial inducers, such as
thio-methylgalactoside (TMG) and isopropyl-$\beta$,
D-thiogalactopyranoside (IPTG), rather than the natural inducer,
allolactose. This difference is critical because artificial inducers
(also known as gratuitous inducers) are not metabolized by the induced
enzyme, whereas the natural inducer is a metabolic intermediate in
lactose degradation, which is catalyzed by the induced enzyme.

Savageau~\cite{Savageau01} found important differences between
induction by IPTG vs. lactose in his theoretical treatment of
bistability in the {\em lac} operon. In Savageau's model, because
production and decay of allolactose are both proportional to the
$\beta$-gal concentration, bistability is forbidden. Expression of
{\em lac} in response to lactose was therefore predicted not to
exhibit bistability. This prediction agreed with the absence of {\em
steady-state} bistability in an experimental study of populations of
{\em E. coli} cells exposed to lactose, described in the Supplementary
Material of Ref.~\cite{Ozbudak04}---in that study, only {\em
transient} bimodal distributions of green fluorescence levels among
cells were observed at some glucose concentrations. It was later noted
that models with operon-independent decay of lactose (e.g., due to
dilution by cell growth) could exhibit
bistability~\cite{vanHoek06}. Several studies using such models found
either a bistable or graded response to lactose, depending on
parameter values or external glucose
levels~\cite{Yildrim03,Yildrim04,vanHoek06,vanHoek07,Santillan07}, and
in agreement with the model of Savageau, a model of van Hoek \&
Hogeweg~\cite{vanHoek06} was explicitly shown to exhibit no
bistability in the absence of operon-independent decay of
allolactose. However, these studies disagree in their assessment of
whether bistability is present~\cite{Yildrim03,Yildrim04,Santillan07}
or absent~\cite{vanHoek06,vanHoek07} in expression of {\em lac} among
{\em E. coli} cells in a natural context.

In addition to predicting whether {\em lac} induction exhibits
bistability, some studies have addressed the question of whether
bistability might enhance or hinder the performance of {\em E. coli}
cells. Both Savageau~\cite{Savageau02} and van Hoek \&
Hogeweg~\cite{vanHoek07} found that bistability increases the time
required to respond to sudden increases in environmental lactose,
which can be a disadvantage in competition for nutrients. These
results argue against the natural relevance of bistability in {\em
lac} expression.  

Another important question that has not yet been addressed is whether
the experimental observations of bistability in Ref.~\cite{Ozbudak04}
are consistent with independent biophysical data that characterize
processes relevant to regulation of {\em lac} expression. Although
phenomenological models were developed to reproduce the steady-state
behavior~\cite{Ozbudak04} and the experimentally characterized
dynamics of switching between stable steady states~\cite{Mettetal06},
these models were not constrained by independent biophysical data. For
example, it is unclear whether the phenomenological models are
consistent with independently measured permease transport kinetics. On
the other hand, studies of bistability using more detailed,
biophysical models of {\em lac} induction were either only partially
constrained~\cite{vanHoek06} or did not consider the response to
artificial inducers~\cite{Yildrim03,Yildrim04,Santillan07}.

Here we analyze bistability in an ordinary differential equation (ODE)
model of {\em lac} induction. We use ODEs because we restrict our
analysis to steady-state behaviors, and because the protein
concentrations in fully induced cells are O($10^4$) per cell (see
Parameter Values section) and have negligible fluctuations. Similar
equations describe induction by lactose or artificial inducers. We
first use the model to gain insight into key determinants of
bistability of {\em lac} expression in response to artificial
inducers, and to understand how characteristics of bistability are
controlled by model parameters. We then use the resulting insight to
tune the parameters of the model to match the bistable behavior
observed by Ozbudak et al.~\cite{Ozbudak04}, and to predict mechanisms
by which bistability might be abolished. Finally, like previous
modeling studies, we use the model to address the question of whether
{\em lac} expression might be bistable in a natural context,
contributing to resolution of what is now a long-standing controversy.

\section{Model}
\label{SEC-MOD}

\begin{figure}[ht]
\begin{center}
\includegraphics[width=5.0in]{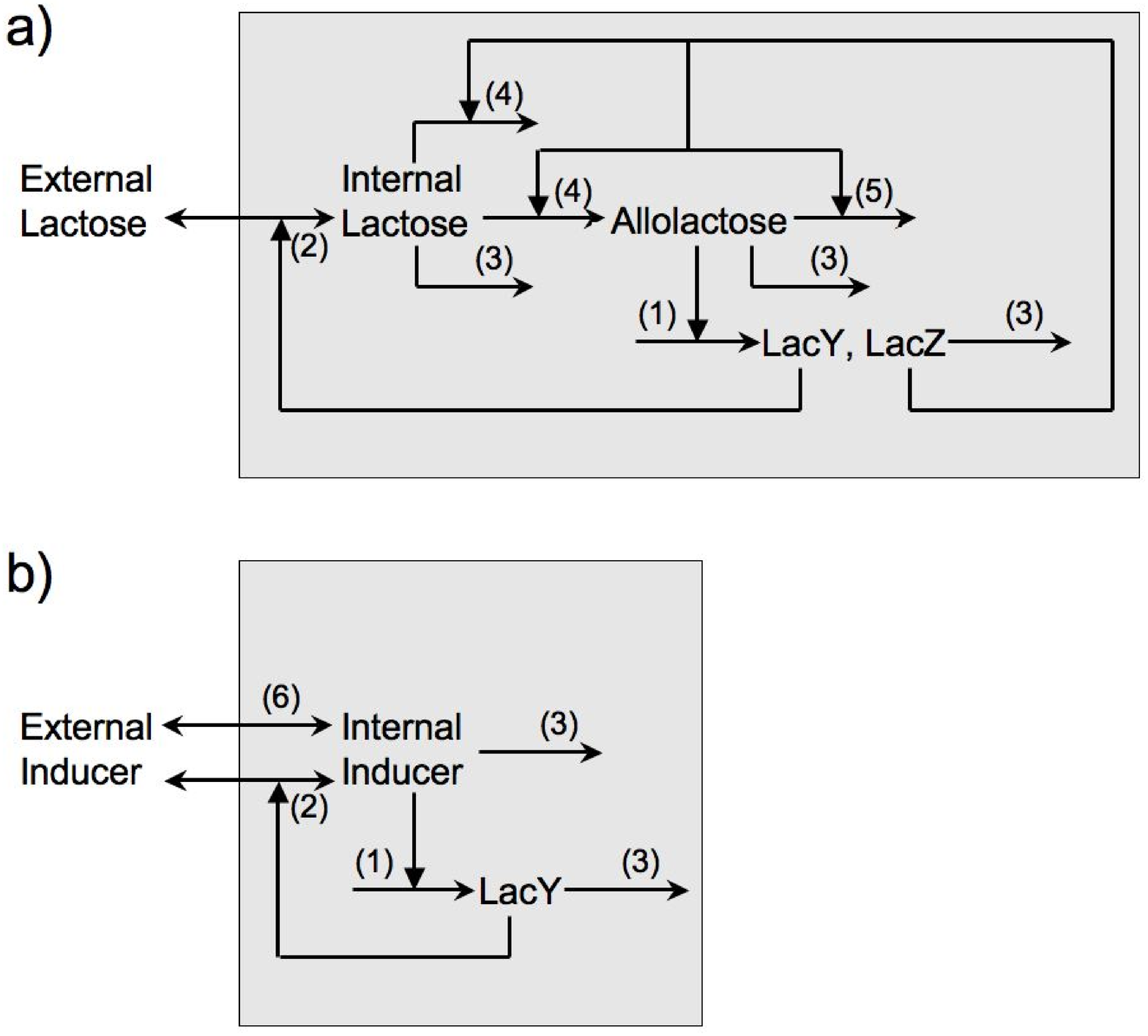}
\end{center}
\caption{Circuitry for models of {\em lac} induction.  a) Model for
induction by lactose (Eqs.~(\ref{allo})), including the following
processes: (1) proportional production of permease (LacY) and
$\beta$-gal (LacZ); (2) permease-mediated transport of lactose; (3)
dilution of intracellular species by cell growth; (4) $\beta$-gal
catalyzed degradation of lactose, producing both the metabolic
intermediate allolactose, and the ultimate products of degradation,
glucose and galactose; and (5) $\beta$-gal catalyzed degradation of
allolactose, producing glucose and galactose. b) Model for induction
by artificial inducers (Eqs.~(\ref{iptg})), including: (1)
proportional production of permease (LacY) and $\beta$-gal (LacZ); (2)
permease-mediated transport of inducer; (3) dilution of intracellular
species by cell growth and (6) passive transport of inducer.}
\label{myfig}
\end{figure}

In our model of {\em lac} induction (Fig.~\ref{myfig}a), the following
set of coupled ordinary differential equations relate the internal
lactose concentration ($l$), allactose concentration ($a$), and
$\beta$-galactosidase concentration ($z$) to the external lactose
concentration ($l^{*}$)
\begin{subequations}
\label{allo}
\begin{eqnarray}
\dot{l} & = &  \alpha z \frac{\left(l^{*} - \phi l\right)}{K_{i} + l+ l^{*}} - \frac{\beta z l}{\left(1+a/K_{m,a}\right)K_{m,l} + l} - \gamma l, \label{allo1}\\ 
\dot{a} & = & \frac{\nu\beta z l}{\left(1+a/K_{m,a}\right)K_{m,l} + l} - \frac{\delta z a}{\left(1+l/K_{m,l}\right)K_{m,a} + a} - \gamma a\ \ \mathrm{and}	 \label{allo2}\\
\dot{z} & = & c\gamma +  \frac{\epsilon\gamma a^{n}}{K_{z}^{n} + a^{n}} - \gamma z. \label{allo3}
\end{eqnarray}
\end{subequations}
In Eqs.~(\ref{allo}), $\alpha$ and $\phi\alpha$ are the rate constants for
permease-dependent lactose import and export, respectively, $K_i$ is
the Michaelis constant for permease-dependent lactose transport
(assumed to be the same for import and export), $\beta$ and $K_{m,l}$
are the rate constant and Michaelis constant for lactose degradation,
$\nu$ is the branching fraction of lactose degradation to allolactose,
$\delta$ and $K_{m,a}$ are the rate constant and Michaelis constant
for allolactose degradation, $\gamma$ is the rate of dilution due to
cell growth, $c\gamma$ and $\epsilon\gamma$ are the basal and
inducible rates of $\beta$-galactosidase production, $K_z$ is the
allactose concentration at half-maximal induction of
$\beta$-galactosidase production, and $n$ is the Hill number for
lactose induction of $\beta$-galactosidase production.

The metabolic fluxes in Eqs.~(\ref{allo}) include the effects of
competition between allolactose and lactose for access to
$\beta$-galactosidase ($\beta$ and $\delta$ terms). Because shuttling
of galactosides across membranes occurs through a single permease
channel \cite{Lolkema91}, we also consider the influence of
competition between external and internal lactose for access to
permease ($\alpha$ and $\phi$ terms); however, as a simplification, we
do not consider transitions among distinct internal states of the
permease \cite{Lolkema91}.

To focus on the operating conditions of
the system that are most relevant to lactose utilization by {\em
E. coli}, we only consider regulation in the absence of glucose. This
focus is appropriate because, in the presence of glucose, {\em lac} is
not essential for growth, and induced $\beta$-galactosidase levels are
low \cite{Magasanik87}. 

Similarly, the model of artificial induction of {\em lac}
(Fig.~\ref{myfig}b) is given by
\begin{subequations}
\label{iptg}
\begin{eqnarray}
\dot{l} & = & \alpha_{0} \left(l^{*} - l \right) + \alpha z\frac{ \left(l^{*} - \phi  l\right)}{K_{i} + l+ l^{*}}  - \gamma l\ \ \mathrm{and}		\label{iptg1} \\
\dot{z} & = & c\gamma + \frac{\epsilon\gamma l^{n}}{K_{z}^{n} + l^{n}} - \gamma z.	\label{iptg2}
\end{eqnarray}
\end{subequations}
In Eqs.~(\ref{iptg}), variables and parameters have the same meaning as
in Eqs.~(\ref{allo}), except $l$ and $l^{*}$ correspond to the level of
internal and external artificial inducer (e.g., IPTG or TMG),
respectively, and $\alpha_0$ is the rate constant for leakage across
the membrane.

In Eqs.~(\ref{allo}) and Eqs.~(\ref{iptg}), protein expression is
lumped with gene expression, and the dependence of promoter activity
on the level of signal (IPTG, TMG, or allolactose) is modeled using a
simple Hill function, which is significantly simpler than other models
\cite{Wong97,Yildrim03,Yildrim04,vanHoek06,Narang07,vanHoek07,Santillan07}. On
the other hand, Eqs.~(\ref{allo}) considers the effects of competition
among substrates in permease transport and metabolic processes, unlike
other models of {\em lac} induction
\cite{Savageau01,Savageau02,Vilar03,Ozbudak04,Wong97,Yildrim03,Yildrim04,vanHoek06,vanHoek07,Santillan07}. Compared
to the model of Savageau \cite{Savageau01,Savageau02},
Eqs.~(\ref{allo}) considers operon-independent decay of allolactose,
without which bistability in response to lactose is impossible
\cite{Savageau01,Savageau02,vanHoek06}, as discussed above. Overall,
Eqs.~(\ref{allo}) and Eqs.~(\ref{iptg}) are less detailed than the
{\em lac} induction models used in Refs. \cite{Wong97},
\cite{Yildrim03}, \cite{Yildrim04}, \cite{vanHoek06},
\cite{vanHoek07}, and \cite{Santillan07}, and are more detailed than
those used in Refs. \cite{Savageau01}, \cite{Savageau02},
\cite{Vilar03}, and \cite{Ozbudak04}, and they therefore constitute
intermediate complexity equations describing {\em lac}
induction. Compared to the simpler models, the intermediate level of
detail provides increased contact between model parameters and
biophysical measurements, and compared to more detailed models, it
facilitates analysis of the equations and interpretation of the
results.

\section{Parameter values}

We used the parameter values and ranges listed in
Table~\ref{Tableparams} to analyze bistability in Eqs.~(\ref{allo})
and Eqs.~(\ref{iptg}). The values in the table were obtained as follows:

\begin{itemize}
\item $\gamma$. We assume the generation time under the conditions in
Ref.~\cite{Ozbudak04} is 30-60 min. We note, however, that this time
might be very different for {\em E. coli} growing under stress in the
gut; this represents a source of uncertainty concerning the biological
relevance of our predictions.

\item $\alpha_0$. We assume that $\alpha_0=0$ except for the case of
IPTG, where we explore a range consistent with that considered in
Ref.~\cite{vanHoek07}.

\item $\alpha$. An approximate range of 1-100~s$^{-1}$ for sugar
transport turnover numbers was obtained from the review by Wright et
al.~\cite{Wright86}. The range is broader than measured values
\cite{Viitanen84} because measurements were made at 25$^\circ$C rather
than at the physiological temperature of 37$^\circ$C in the host
environment of the gut that we are focusing on here, and at which
measurements in Ref. \cite{Ozbudak04} were performed. The nominal
value of $1000\ \mathrm{min}^{-1}$ was estimated from
Ref.~\cite{Viitanen84} assuming the production rate of permease is the
same as that of functional $\beta$-gal. Because permease is a monomer
while $\beta$-gal is a tetramer, this assumption entails a four-fold
smaller production rate for permease. This seems possible, as (1)
galactoside acetyltransferase (GATase) monomer synthesis is eight-fold
smaller than $\beta$-gal monomer synthesis; (2) due to incomplete
operon transcription and the order of genes in the operon ({\em
lacZYA}), the amount of mRNA transcribed from the GATase gene ({\em
lacA}) and permease gene ({\em lacY}) is smaller than that from the
$\beta$-gal gene ({\em lacZ}); (3) there is some evidence that
permease is made in smaller amounts than $\beta$-gal \cite{Kennedy70}.

\item $\phi$. We assume no export flux through permease in the
artificial induction model, and then examine the consequences of
introducing such a flux on bistability. Guided by
Ref.~\cite{Lolkema91}, for the lactose model, we use a nominal efflux
rate constant ($\phi\alpha$) of half the value of the influx rate
constant $\alpha$, and allow the value to decrease in the search for
bistable conditions.

\item $K_i$. For simplicity, we assume the same Michaelis constant for
permease import and export--a nominal value of $0.5$~mM was obtained
from Ref.~\cite{Viitanen84}. The range was applied as per $\alpha$,
and encompasses measured values \cite{Wright81,Page84,Viitanen84}.

\item $\beta$. A total lactose turnover number for
$\beta$-galactosidase of $2.85\times 10^4$~min$^{-1}$ is estimated
from a measured value of $V_{max}=61.3~\mu$mol~min$^{-1}$~mg$^{-1}$ in
Ref.~\cite{Huber76}. This estimate is an order of magnitude greater
than the value $3.6\times 10^3$~min$^{-1}$ given in
Ref.~\cite{Martinez91}, but the two estimates agree closely when one
considers that $\beta$-gal converts about half of its lactose
substrate to glucose and galactose, rather than allolactose, and that
the enzyme is composed of four monomeric catalytic subunits. The
estimate given in Ref.~\cite{Martinez91} is appropriate for total
turnover of lactose on a per monomer basis. Like for $\alpha$, because
measurements were performed at 30$^\circ$C, we consider a range of
values ten times lower to ten times higher than the nominal value.

\item $K_{m,l}$.  The nominal value was obtained directly from
Ref.~\cite{Huber75}. As for $\beta$, because of temperature
considerations, we use a range from ten times lower to ten times
higher than the nominal value.

\item $\nu$. The value $\nu=0.468$ was calculated from the total rate
of $\beta$-gal degradation of lactose and the partial flux from
lactose to allolactose reported in Ref.~\cite{Huber76}. We take it to
be a constant because the ratio of reaction products was found to be
insensitive to temperature changes between $30^\circ$C and $0^\circ$C.

\item $\delta$. An allolactose turnover number for $\beta$-gal of
$2.3\times 10^4$~min$^{-1}$ is estimated from a measured value of
$V_{max}= 49.6~\mu$~mol~min$^{-1}$~mg$^{-1}$ in
Ref.~\cite{Huber75}. As for $\beta$, because of temperature
considerations, we use a range from ten times lower to ten times
higher than the nominal value.

\item $K_{m,a}$.  The nominal value was obtained directly from
Ref.~\cite{Huber75}. As for $\beta$, because of temperature
considerations, we use a range from ten times lower to ten times
higher than the nominal value.

\item $\epsilon$. Using a production rate of 5 $\beta$-gal tetramers
per cell per second for a 48 min generation time \cite{Kennell77},
14,400 molecules are produced during a generation at full
induction--this is the number of molecules in the cell after doubling
(supporting our choice of a noiseless model). Assuming a 1
$\mu\mathrm{m}^3$ mean cell volume \cite{Kubitschek86} and linear
volume increase in time \cite{Kubitschek90}, the volume after doubling
is approximately 0.7 $\mu\mathrm{m}^3$, leading to a concentration of
34,286 $\mathrm{nM}$.

\item $c$. This value is derived from $\epsilon$, assuming a 1000-fold
increase in $\beta$-galactosidase levels upon induction
\cite{Beckwith87}.

\item $K_z$ and $n$. These values are estimated from IPTG induction
data in permease knockout cells both from Ref.~\cite{Sadler65},
Fig. 15 and from data compiled in Ref.~\cite{Yagil71}, Figs. 1 and
2. The nominal value $n=2$ was estimated from the slopes of the curves
in the figures, and $K_z$ was determined by estimating from the
figures the concentration of IPTG at half-maximal induction. The
nominal value of $10^5\ \mathrm{nM}$ was estimated from data compiled
in Ref.~\cite{Yagil71}. To determine the range, an approximate lower
value of $10^4\ \mathrm{nM}$ was obtained from Ref.~\cite{Sadler65},
and we allowed for an upper value of $10^6\ \mathrm{nM}$ to account
for potential differences between induction by IPTG and TMG or
lactose.

\end{itemize}

%
%
\begin{singlespace}
\begin{table}[htd]		
\caption{Parameter values.}
\begin{center}		
\begin{tabular}{|c|l|c|c|}
\hline
Param & Description & Nominal & Range \\ \hline
	$\gamma$	& growth rate	& -- &	 $0.0116\ \mathrm{min}^{-1}$ -- $0.0231\ \mathrm{min}^{-1}$						\\ \hline
	$\alpha_{0}$ 	& passive transport	&0 &	$0$ -- $1.35\ \mathrm{min}^{-1}$								\\
&rate constant& &\\ \hline
	$\alpha$ 	& permease import	&$600\ \mathrm{min}^{-1}$ & 	$6\times 10^{1}\ \mathrm{min}^{-1}$ -- \\
& turnover number & &$6\times 10^{3}\ \mathrm{min}^{-1}$\\ \hline
	$\phi$		& ratio of permease	&0 (artificial inducers)&	 0 -- $0.5$							\\
&export to import& or 0.5 (lactose)&\\ 
&turnover numbers& &\\ \hline
	$K_{i}$		& permease Michaelis	& $5\times 10^5\ \mathrm{nM}$&	 $5 \times 10^{4}\ \mathrm{nM}$ --\\
&constant& &$5\times 10^6\ \mathrm{nM}$ \\ \hline
	$\beta$		& $\beta$-gal lactose	&$2.85\times 10^4\mathrm{min}^{-1}$ &	 $2.85\times 10^{3}\ \mathrm{min}^{-1}$ -- \\
&turnover number& &$2.85\times 10^{5}\ \mathrm{min}^{-1}$\\ \hline
	$\nu$		& lactose$\rightarrow$ allolactose 	&0.468 &--	 		\\
&$\beta$-gal branching& &\\
&fraction& &\\ \hline
	$K_{m,l}$	& $\beta$-gal lactose		& 2.53 mM&	 $0.253\ \mathrm{mM}$ -- $25.3\ \mathrm{mM}$				\\
&Michaelis constant & & \\ \hline
	$\delta$	& $\beta$-gal allolactose		& $2.30\times 10^4\mathrm{min}^{-1}$&	 $2.30\times 10^{3}\ \mathrm{min}^{-1}$ -- \\
&turnover number& &$2.30\times 10^{5}\ \mathrm{min}^{-1}$\\ \hline
	$K_{m,a}$	& $\beta$-gal allolactose	& 1.2 mM&	 $0.12\ \mathrm{mM}$ --$12.0\ \mathrm{mM}$ \\
&Michaelis constant& &\\ \hline
	$\epsilon$	& fully induced 	& 	 $34285\ \mathrm{nM}$&-- 									\\
& $\beta$-gal level & & \\ \hline
	$c$		& basal $\beta$-gal
 level		& 	 $34.3\ \mathrm{nM}$ &--									\\ \hline
	$K_{z}$		& signal level at	& $10^5$ nM&	 $10^{4}\ \mathrm{nM}$ -- $10^6\ \mathrm{nM}$									\\
&half-maximal & & \\ 
&{\em lac} induction& &\\ \hline
	$n$		& Hill number for		& 	 $2$	&--												\\	
&signal-dependent& &\\
&{\em lac} induction& &\\ \hline
\end{tabular}
\end{center}
\label{Tableparams}
\end{table}
\end{singlespace}
%
%

\section{Results}
\label{SEC-ANAL}

\begin{figure}[ht]
\begin{center}
\includegraphics[width=5.0in]{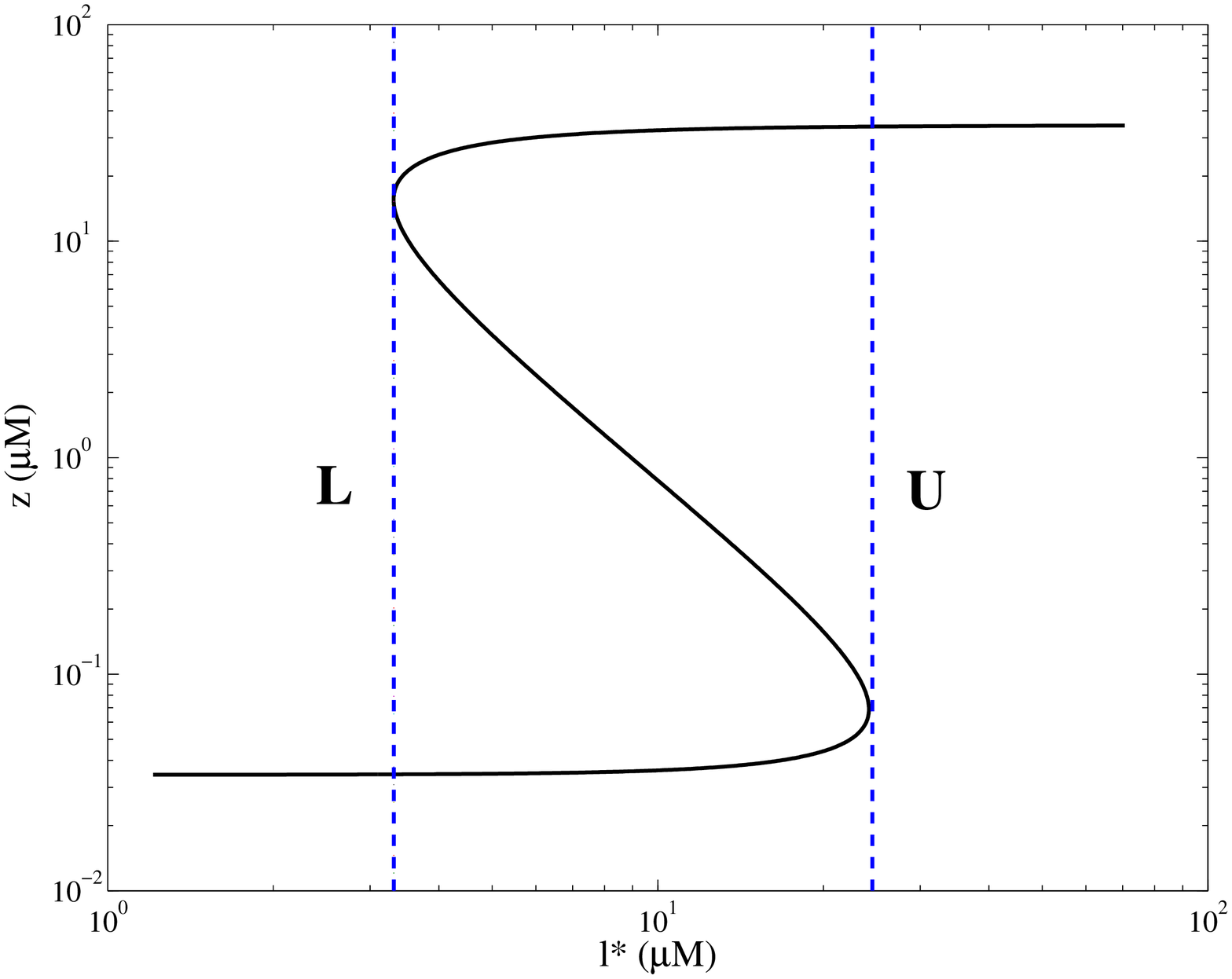}
\end{center}
\caption{An example of a system from Eqs.~(\ref{iptg}) with the upper
($U$) and lower ($L$) turning points consistent with the results in
\cite{Ozbudak04}.  The parameter values are $\gamma = .0231\
\mathrm{min}^{-1}$, $\alpha = 60\ \mathrm{min}^{-1}$, $K_{z} =
123,285\ \mathrm{nM}$ and $K_{i} = 1,077,217\ \mathrm{nM}$.}
\label{OzbTPs}
\end{figure}

%
%
We first used Eqs.~(\ref{iptg}) to determine how parameter values
control bistability in the steady-state response of {\em lac}
expression to artificial inducers. To detect and characterize
bistability for a given set of parameter values, we solved for $z(l)$
and $l^{*}(l)$ as rational functions of $l$. Bistability in {\em lac}
expression exists when the line describing steady-state levels of $z$
vs. $l^{*}$ adopts a characteristic ``S'' shape, as shown in
Fig.~\ref{OzbTPs}. Within the bistable range of $l^{*}$, the highest
and lowest levels of $z$ are stable steady-state solutions and the
intermediate level of $z$ is an unstable steady-state solution of
Eqs.~(\ref{allo}). The bistable range is defined by the lower
($l^{*}=L$) and upper ($l^{*}=U$) turning points, as illustrated in
Fig.~\ref{OzbTPs}. An analogous signature of bistability can be seen
in examining steady-state levels of $l$ vs. $l^{*}$ (not shown). For a
model with given parameter values, $L$ and $U$ can be located by
finding the roots of either $d l^* / d z$ or $d l^{*} / d l$ using an
eigenvalue solver.

We analyzed Eqs.~(\ref{iptg}) for systems with sets of
parameter values drawn from the ranges in Table~\ref{Tableparams},
taking $\alpha_0=0$, $\phi=0$, and $n=2$. Sets of 100 values each for
$K_i$ and $K_z$ were obtained using logarithmically even sampling over
their allowed ranges. Because the steady-state solutions of
Eqs.~(\ref{iptg}) only depend on $\alpha$ and $\gamma$ through the
ratio $\alpha/\gamma$, rather than sampling $\alpha$ and $\gamma$
individually, we obtained 100 values of $\alpha/\gamma$ using
logarithmically even sampling between the upper and lower bound
computed from Table~\ref{Tableparams}. This sampling scheme yielded
$100\times 100\times 100 = 10^6$ systems with different
values of $(\alpha/\gamma,K_i,K_z)$.

We found that all $10^6$ systems exhibited some degree of bistability
in response to induction by artificial inducers. The dependence of the
range of bistability on model parameters was further analyzed using
two measures that we introduce here: the ratio $U/L$, and the product
$UL$. We used these measures to estimate the percentage of systems for
which bistability might be observable in an experiment like that in
Ref.~\cite{Ozbudak04}. By inspecting the measurement errors in
Ref.~\cite{Ozbudak04}, we estimate that systems with $U/L>1.1$ and
$UL>0.01\ \mu\mathrm{M}^2$ exhibit bistability that is favorable for
experimental observation (i.e., difficult to detect), and that systems
with either $U/L<1.1$ or $UL<0.01\ \mu\mathrm{M}^2$ exhibit
bistability that is unfavorable for experimental observation. Among
systems with parameter values sampled as described above, by these
criteria, we predict that experimental observation of bistability
would be favorable for $65\%$ of systems, and unfavorable for $35\%$
of systems.

\begin{figure}[ht]
\begin{center}
\includegraphics[width=5.0in]{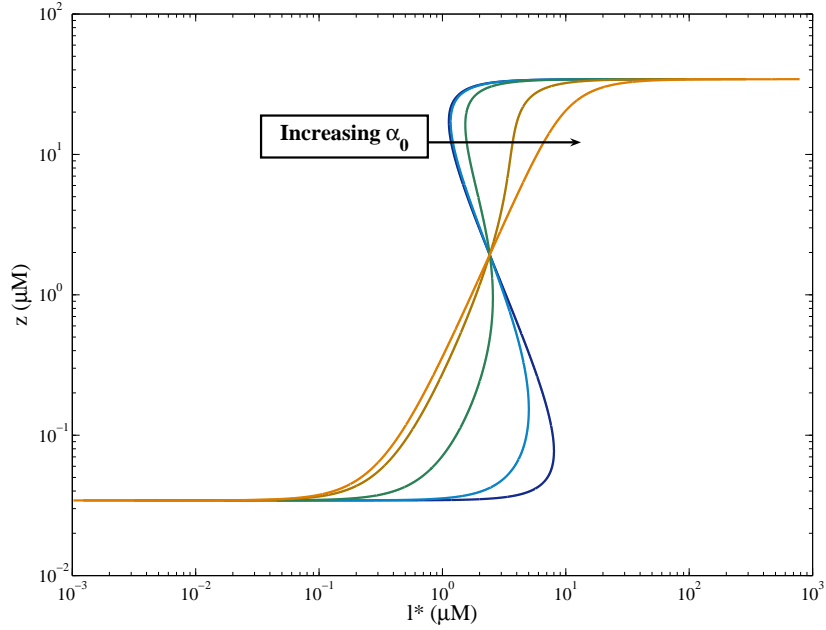}
\end{center}
\caption{Effects of variations in the $\alpha_{0} > 0$ parameter on an
artificially induced system with $\phi = 0\ \mathrm{min}^{-1}$ and
$\alpha_{0} = 10^{-k}\ \mathrm{min}^{-1}$, $k = 0,\ldots, 4$. The
other parameters are given by $n = 2,\ \gamma = .0231\
\mathrm{min}^{-1},\ \epsilon = 34286\ \mathrm{nM},\ c = 34.3\
\mathrm{nM},\ K_{i} = 5 \times 10^{6}\ \mathrm{nM},\ K_{z} = 10^4\
\mathrm{nM}$ and $\alpha = 60\ \mathrm{min}^{-1}$.}
\label{alpha}
\end{figure}

\begin{figure}[ht]
\begin{center}
\includegraphics[width=5.0in]{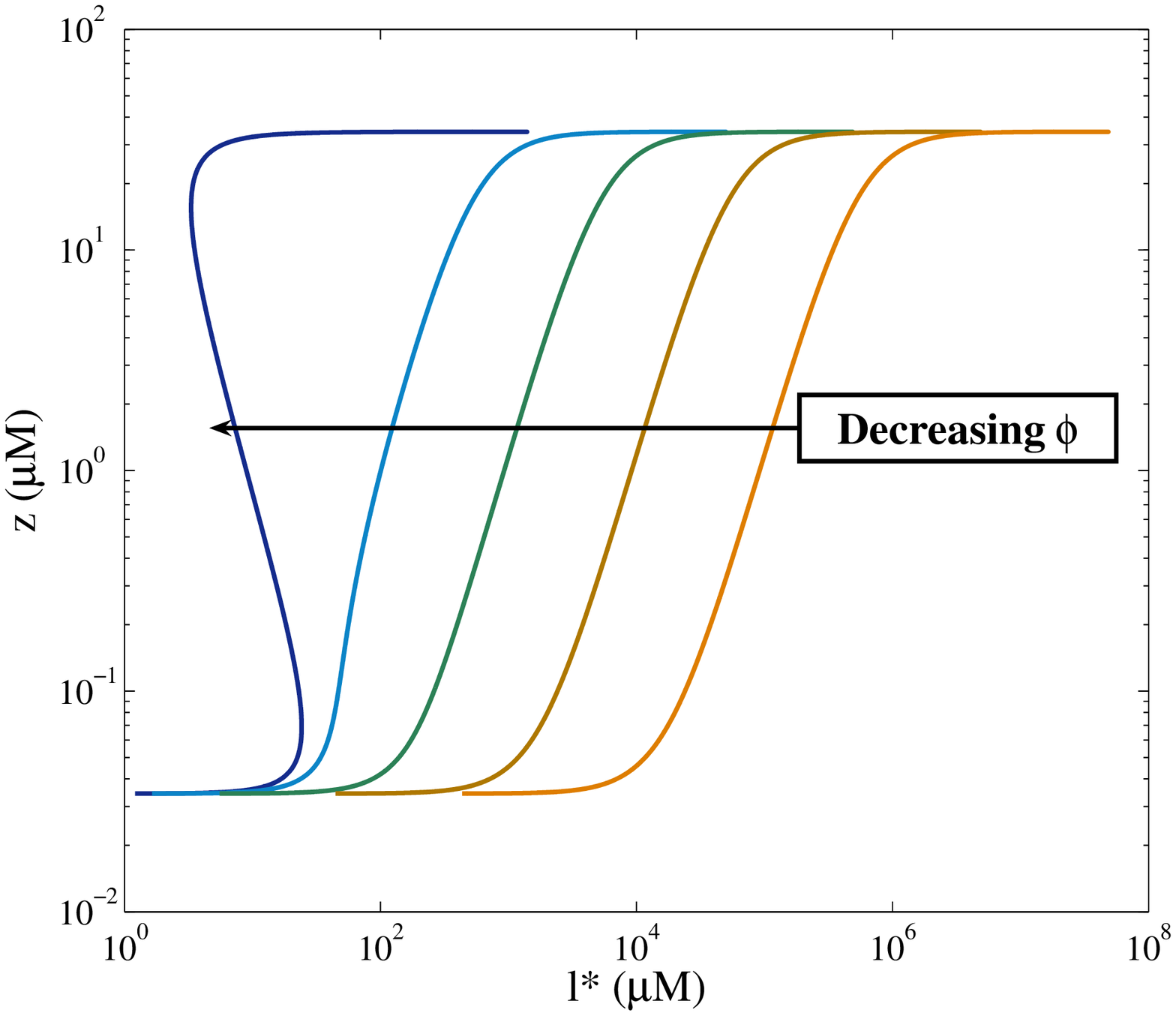}
\end{center}
\caption{Effects of variations in the $\phi > 0$ parameter on an
artificially induced system with $\alpha_{0} = 10^{-4}\
\mathrm{min}^{-1}$ and $\phi = 0$ and $10^{-k}\ \mathrm{min}^{-1}$, $k
= 1,\ldots,4$.  All of the scales are in $\mu$M.  The other parameter
values are as in Figure~\ref{alpha}.}
\label{phi}
\end{figure}

Increasing either $\alpha_{0}$ or $\phi$ above zero tends to reduce or
abolish bistability in artificially induced systems.  As
$\alpha_{0}$ is increased (Fig.~\ref{alpha}), first $U$ begins
shifting to lower values of $l^{*}$, then $L$ begins shifting to
higher values of $l^{*}$, leading to an asymptotic behavior in which
bistability is abolished.  Like changes in $\alpha_0$, as $\phi$ is
increased (Fig.~\ref{phi}), $L$ shifts to higher values of $l^{*}$;
however, by contrast, $U$ does not initially show a significant
change.  As $\phi$ is increased further, the entire induction curve
begins to shift to higher levels of $l^{*}$.

\begin{figure}[htd]
\begin{center}
\includegraphics[width=5.0in]{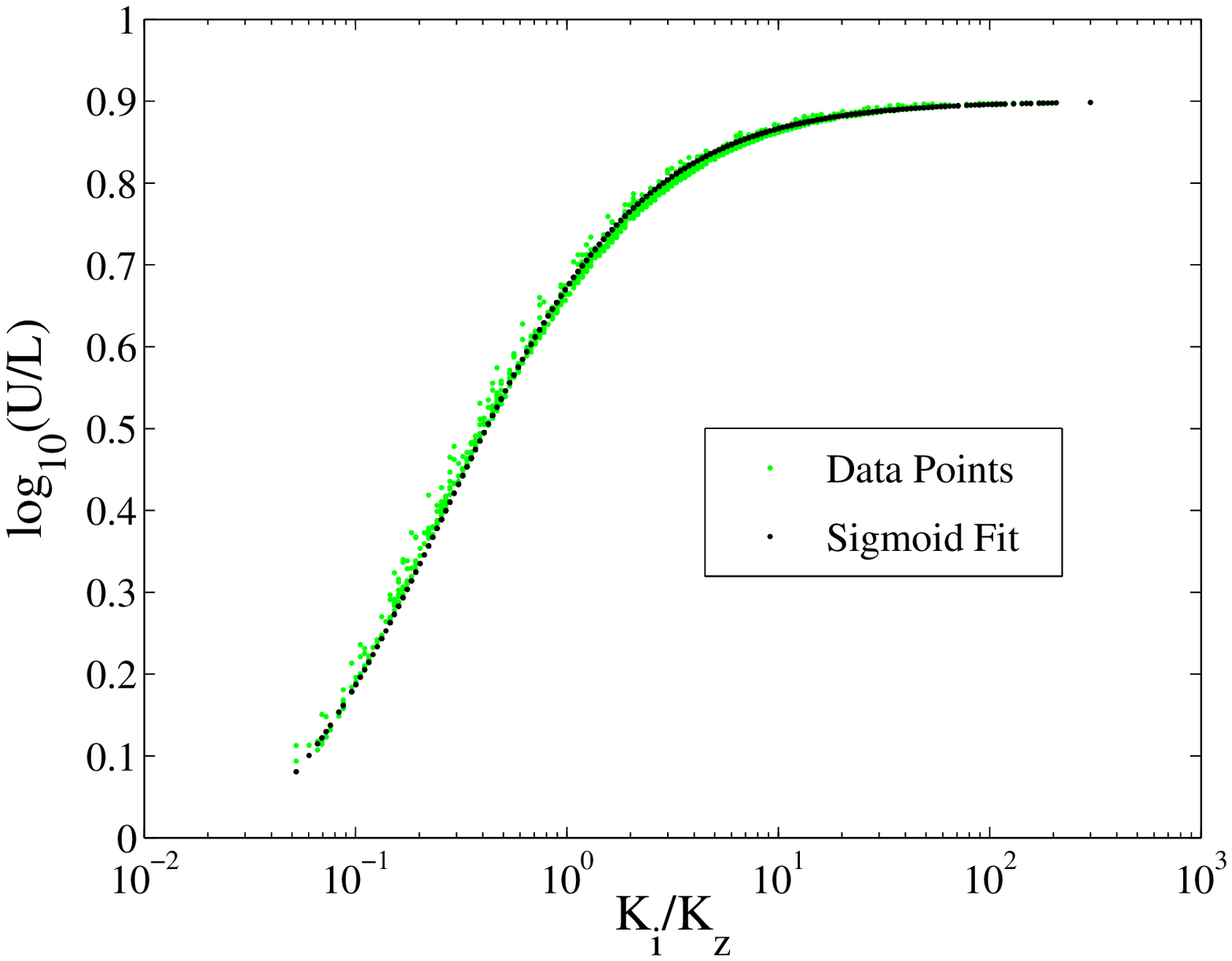}
\end{center}
\caption{Modeling the width as a sigmoid function of $K_{i}/K_{z}$.  Only a sample of data points are shown.}
\label{WidthKiKz}
\end{figure}

\begin{figure}[htd]
\begin{center}
\includegraphics[width=5.0in]{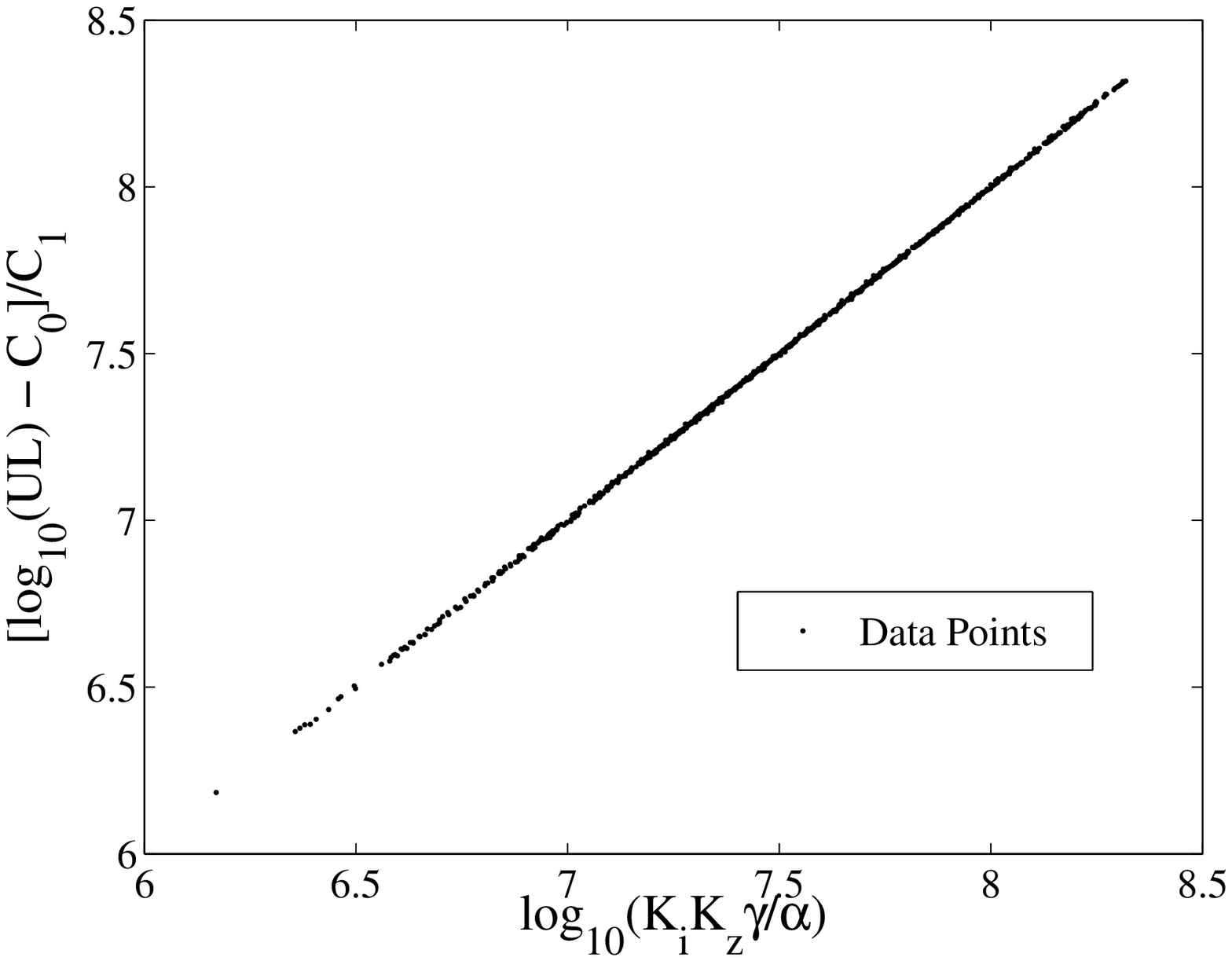}
\end{center}
\caption{Using the $C_{0}$ and $C_{1}$ in Figure~\ref{CenterCoeffs}
results in a linear relation between the center $\log_{10}(UL)$ and
$\log_{10}(K_{i}K_{z}\gamma/\alpha)$.}
\label{CenterReg}
\end{figure}

\begin{figure}[htd]
\begin{center}
\includegraphics[width=5.0in]{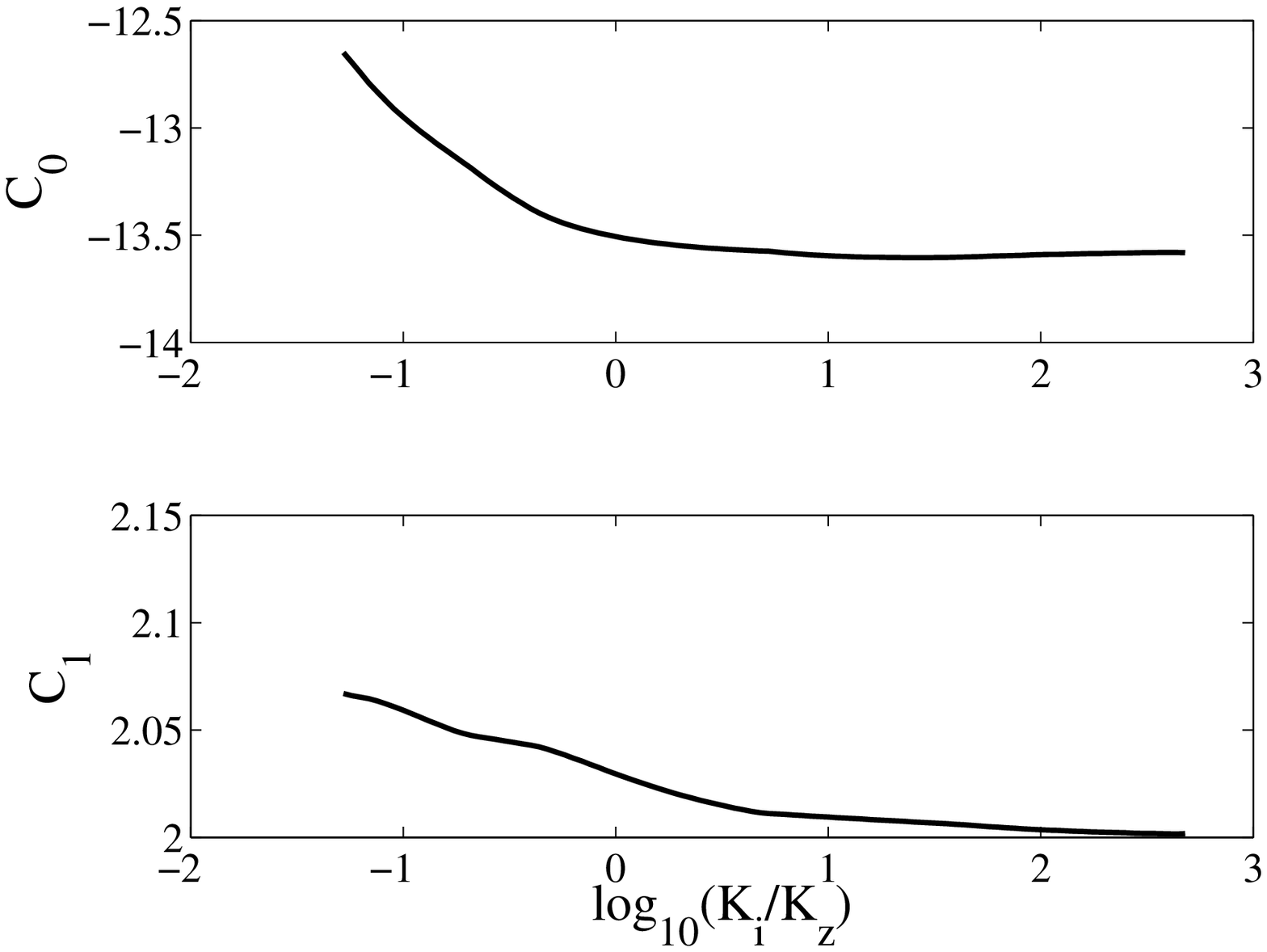}
\end{center}
\caption{Dependence of the coefficients in the regression model of
Eq.~(\ref{centereq}) on $X=K_i/Kz$.}
\label{CenterCoeffs}
\end{figure}

To compare Eqs.~(\ref{iptg}) to the data in Ref.~\cite{Ozbudak04}, we
first selected a subset of systems for which the bistable region is in
the same neighborhood as that in Ref.~\cite{Ozbudak04}: from $3\ \mu
\mathrm{M}$ to $30\ \mu \mathrm{M}$ TMG. Considering this range, out
of the $10^6$ systems sampled, we selected 187,108 systems for which
$L>1\ \mu$M and $U<\ 100\ \mu$M for further analysis. Interestingly, we
found that all of these systems collapse to a single curve when
displayed in the space of $\log_{10}(U/L)$ vs. $\log (K_{i}/K_{z})$
(Fig.~\ref{WidthKiKz}), indicating that $U/L$ can be precisely tuned
using the parameter $X = K_{i}/K_{z}$.  As shown in
Fig.~\ref{WidthKiKz}, the dependence was accurately modeled using the
equation
\begin{eqnarray}
\label{widtheq}
\log_{10} (U/L) & \approx &  \frac{\ (K_{i}/K_{z})^{.93}}{(K_{i}/K_{z})^{.93} + (.27)^{.93}} - \frac{1}{10} \geq 0.
\end{eqnarray}
Next, we found that, at a given value of $X=K_{i}/K_{z}$, without
changing the value of $U/L$, $UL$ could be tuned precisely using the
parameter $Y = K_{i}K_{z}\gamma/\alpha$. As shown in
Fig.~\ref{CenterReg}, this dependence was accurately modeled using the
equation
\begin{eqnarray}
\label{centereq}
\log_{10} (UL) & = & C_{0}(X) + C_{1}(X) \log_{10}(Y).
\end{eqnarray}
Figure~\ref{CenterCoeffs} shows the $X$-dependence of the parameters
$C_0(X)$ and $C_1(X)$, obtained numerically using systems with similar
values of $X$.  For the range of systems considered here, we found
that $C_{0}(X)$ could be fit using a third order polynomial in
$\log_{10}(X)$, and that $C_{1}(X)$ could be taken as a constant.

The above phenomenological results provide a prescription for tuning
the range of bistability exhibited by an artificially induced
system. First, the value of $U/L$ can be specified by choosing a value
of the parameter $X=K_{i}/K_{z}$ using Eq.~(\ref{widtheq}). Then,
using this value of $X$, the value of $UL$ can be specified by
choosing a value of the parameter $Y=K_z K_i \alpha/\gamma$ using
Eq.~(\ref{centereq}) and the empirically determined $C_0(X)$ and
$C_1(X)$ (Fig.~\ref{CenterCoeffs}). We used this prescription to
obtain a family of systems that are consistent with the parameter
values in Table~\ref{Tableparams} and that exhibit a range of
bistability consistent with that observed in Ref. \cite{Ozbudak04},
with $\log_{10}(U/L) \approx .86$ and $\log_{10}(UL) \approx 1.92$. An
example of the steady-state behavior of one such system is illustrated
in Figure~\ref{OzbTPs}.

\begin{figure}[ht]
\begin{center}
\includegraphics[width=5.0in]{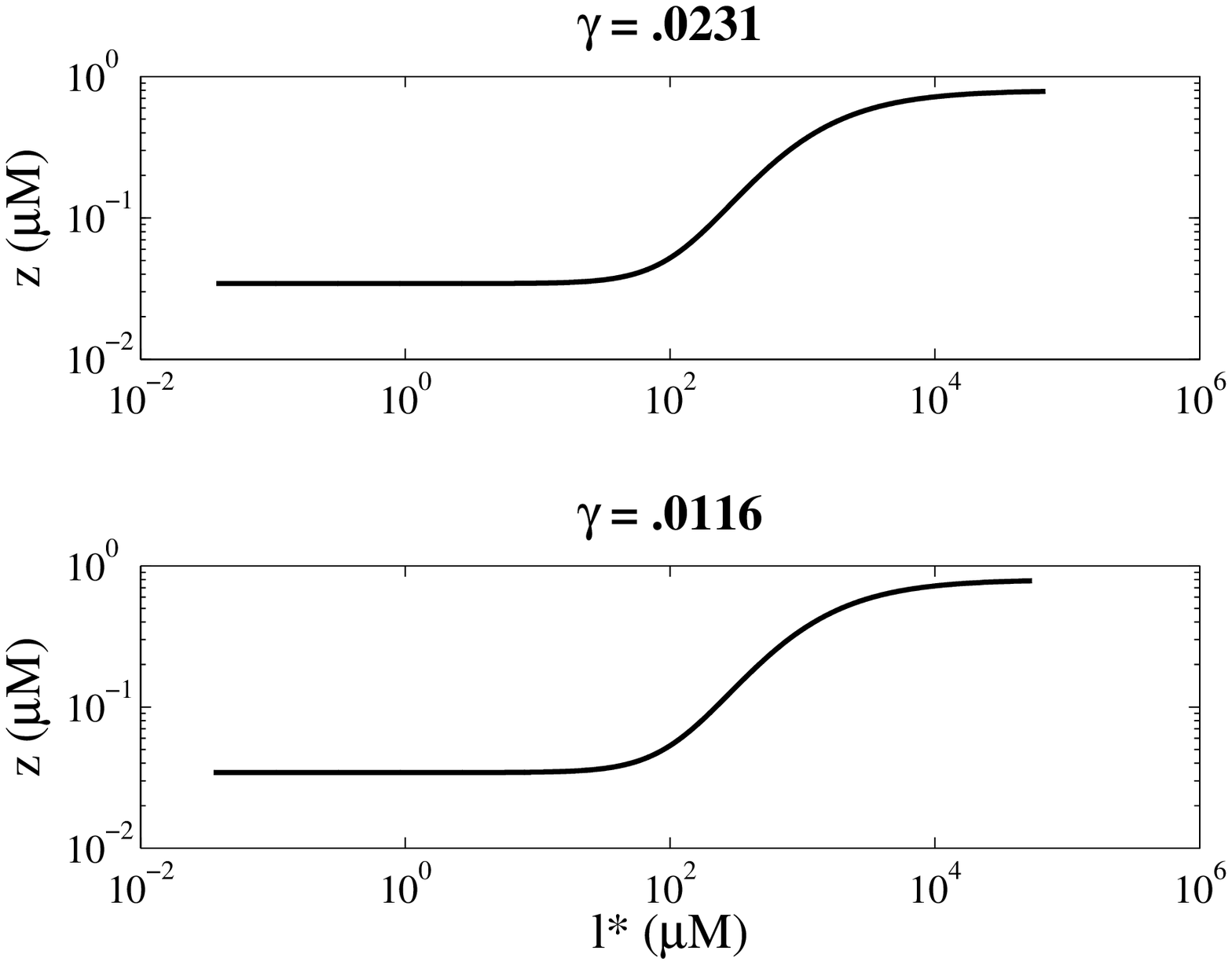}
\end{center}
\caption{Allolactose system with all of the parameters given by their
nominal values in Table~\ref{Tableparams}.}
\label{allosys}
\end{figure}

\begin{figure}[ht]
\begin{center}
\includegraphics[width=5.0in]{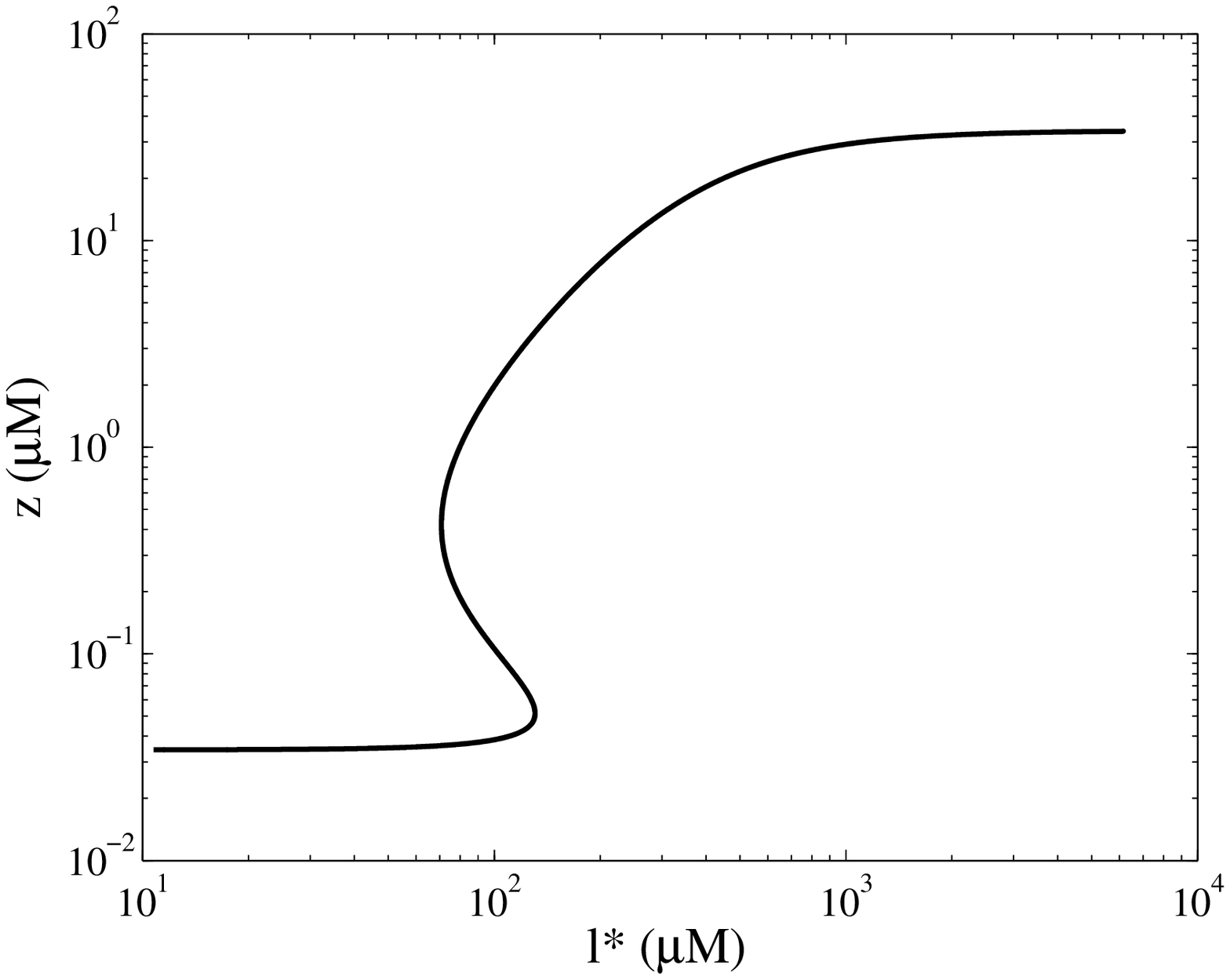}
\end{center}
\caption{Bistability in the $\phi = 0$ lactose-induced system with
$\alpha$, $\beta$, $\delta$ and $K_{z}$ at their lowest values in
Table~\ref{Tableparams} and $\gamma$, $K_{m,l}$, $K_{m,a}$ and $K_{i}$
at their highest values. This is the system that exhibits the largest
values of $U/L$ and $UL$ within the allowed ranges of parameter
values.}
\label{Biallosys}
\end{figure}

We used similar methods to analyze Eqs.~(\ref{allo}) which describe
induction by lactose. No bistability was present in the system with
nominal parameter values from Table~\ref{Tableparams}
(Fig.~\ref{allosys}) with $\phi=0.5$, which is consistent with the
theory of Savageau~\cite{Savageau01} and the Supplementary Material of
Ref.~\cite{Ozbudak04}. However, guided by the results for artificial
inducers in Fig.~\ref{phi}, we examined systems with
$\phi=0$. Although the system with nominal parameter values and
$\phi=0$ did not exhibit bistability, other systems that have
parameter values consistent with the ranges in Table~\ref{Tableparams}
did exhibit bistability. We then located the system that exhibits the
largest values of $U/L$ and $UL$; for this case, $\alpha$, $\beta$,
$\delta$ and $K_{z}$ assume their lowest values in
Table~\ref{Tableparams} while $\gamma$, $K_{m,l}$, $K_{m,a}$ and
$K_{i}$ assume their highest values (Fig.~\ref{Biallosys}). The curve
in Fig.~\ref{Biallosys} illustrating bistability characteristics for
this system closely resembles a similar curve shown in van Hoek \&
Hogeweg~\cite{vanHoek06}, Fig. 2B. Thus, although our model is less
detailed than theirs, it can exhibit comparable steady-state behavior.

\begin{figure}[ht]
\begin{center}
\includegraphics[width=5.0in]{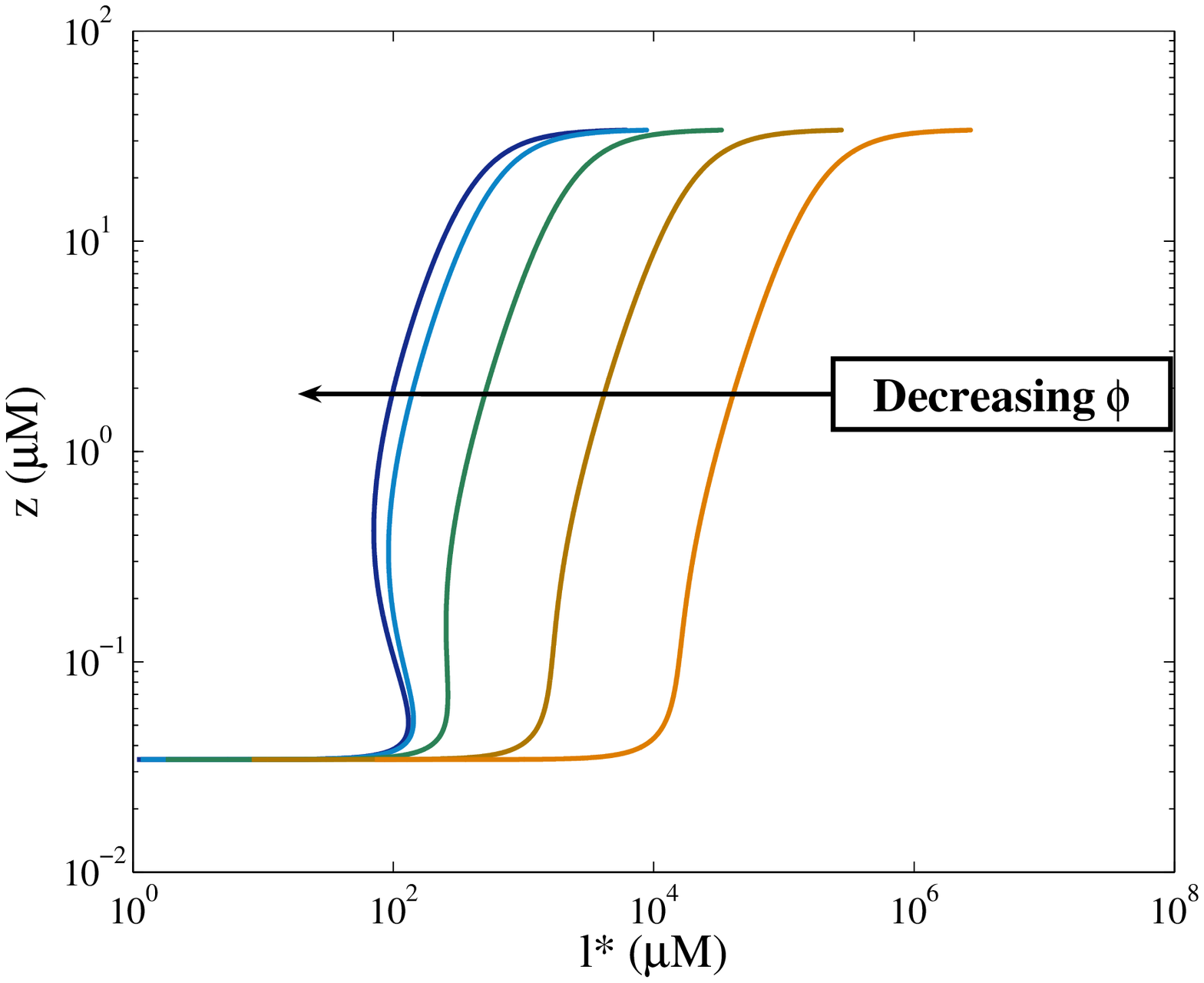}
\end{center}
\caption{This is the same system as in Figure~\ref{Biallosys} with $\phi = 0$ and $10^{-k}\ \mathrm{min}^{-1}$, $k = 1,\ldots,4$.}
\label{Allophi}
\end{figure}

To estimate the distribution of systems exhibiting the different
qualitative behaviors, as for the case of artificial inducers, we
analyzed $10^4$ systems with randomly sampled parameter values, all
with $\phi = 0$.  We predict $99.87\%$ of these systems to exhibit no
bistability, $0.05\%$ to exhibit bistability favorable for observation
($U/L > 1.1$ and $UL > 0.01\ \mu \mathrm{M}^2$), and $0.08\%$ to
exhibit bistability that is unfavorable for observation ($U/L < 1.1$
or $UL < 0.01\ \mu \mathrm{M}^2$). However, as observed for
Eqs.~(\ref{iptg}), increasing $\phi$ to even a small fraction of its
nominal value rapidly abolishes bistability for all combinations of
other parameter values in Eqs.~(\ref{allo}) (Fig.~\ref{Allophi}).

\section{Conclusions}
\label{SEC-CONC}

For the equations describing induction by artificial inducers, we
found that the range of external inducer concentrations over which
systems exhibit bistability is precisely controllable by two rational
combinations of model parameters. By adjusting these parameters, we
were able to demonstrate agreement with the bistable range for TMG
induction from Ref.~\cite{Ozbudak04}. However, in achieving this
agreement, we assumed that permease-dependent efflux of artificial
inducers is negligible ($\phi=0$). We have not found independent
biophysical data to constrain this parameter for artificial inducers,
and therefore predict that it has a value much less than the value of
roughly $0.5$ that has been measured for lactose.

To achieve agreement with the bistable range of roughly $3\ \mu
\mathrm{M}$ to $30\ \mu \mathrm{M}$ in Ref.~\cite{Ozbudak04}, $c$ and
$\epsilon$ in Eqs.~(\ref{iptg}) were tuned to exhibit a 1000-fold
induction of protein expression. While this value is reasonable based
on previous studies, it does disagree with the roughly 100-fold
induction of GFP expression reported in Ref.~\cite{Ozbudak04}. We did
analyze systems with alternative values of $c$ and $\epsilon$ that
yield 100-fold induction; however, none of them exhibited bistable
ranges that agree with the range reported in
Ref.~\cite{Ozbudak04}. Further studies will be required to understand
why Eqs.~(\ref{iptg}) does not simultaneously agree with both the
bistable range and maximal induction of the experimental $lac::gfp$
reporter system. In addition to model refinement, it would be fruitful
to seek systematic differences between expression from chromosomal
{\em lac} and the plasmid-based $lac::gfp$ reporter system used in
Ref.~\cite{Ozbudak04}.

The lack of bistability observed for induction by lactose agrees with
modeling studies concluding that bistabity in {\em lac} expression is
irrelevant to {\em E. coli} in a natural
context~\cite{Savageau01,Savageau02,vanHoek06,vanHoek07}. Thus,
although bistable behavior in {\em lac} is now
well-documented~\cite{Novick57,Cohn59c,Ozbudak04}, because it has only
been experimentally observed using artificial inducers, its relevance
within the natural context of {\em E. coli} is doubtful. Indeed, it is
surprising that the {\em lac} operon has been considered to be a
paradigm of bistability in gene regulation, considering the gaps in
understanding that remain after so many careful experimental and
theoretical studies.

The present results predict that bistable behavior can be promoted by
(1) hindering the kinetics of permease transport ($\alpha$, $K_i$) and
$\beta$-gal catalysis ($\beta$, $\delta$, $K_{m,l}$, $K_{m,a}$); (2)
lowering the required level of allolactose for half-maximal {\em lac}
expression ($K_z$); and (3) accelerating cell growth ($\gamma$). These
predictions suggest genetic targets for engineering {\em E. coli}
strains that exhibit a clear signature of bistability. Experiments to
compare the behavior of such strains with wild-type cells would help
to clarify whether bistability in {\em lac} expression is relevant in
a natural context.


\newpage

\begin{acknowledgments}
\label{SEC-ACK}

Supported by the US Department of Energy through contract
DE-AC52-06NA25396, and grant GM 80216 from the National Institutes of
Health. The collaboration was facilitated by the First q-bio Summer
School on Cellular Information Processing, which was sponsored by the
New Mexico Consortium's Institute for Advanced Studies, and the Center
for Nonlinear Studies at Los Alamos National Laboratory.

\end{acknowledgments}


\newpage
\bibliography{Lac}

\end{document}